\documentclass[12pt]{article}
\usepackage{amsmath,amssymb}
\usepackage{graphicx}
\usepackage{epstopdf}
\usepackage[pdftex]{hyperref}
\DeclareGraphicsExtensions{.eps,.bmp,.wmf,.jpeg,.pdf}
\numberwithin{equation}{section}
\def\be{\begin{equation}}
\def\ee{\end{equation}}

\def\bea{\begin{eqnarray}}
\def\eea{\end{eqnarray}}

\title{Non-minimal kinetic coupling and Chaplygin gas cosmology}

\author{L. N. Granda$^{1,2}$ \thanks{ngranda@univalle.edu.co, ngranda@um.es} ,  E. Torrente-Luj\'an$^{2,}$ \thanks{torrente@cern.ch, etl@um.es} and J. J. Fernandez-Melgarejo$^{2,}$ \thanks{jj.fernandezmelgarejo@um.es} \\{\small\it $^{1}$Departamento de F\'{i}sica, Universidad del Valle, 25360 Cali Colombia}\\
{\small\it $^{2}$Departamento de F\'{i}sica, Universidad de Murcia}, \\ {\small\it Campus Espinardo, E-30100 Murcia Spain}} 
\date{}
\begin{document}
\maketitle

\begin{abstract}
\noindent In the frame of the scalar field model with non minimal kinetic coupling to gravity, we study the cosmological solutions of the Chaplygin gas model of dark energy. By appropriately restricting the potential, we found the scalar field, the potential and coupling giving rise to the Chaplygin gas solution. Extensions to the generalized and modified Chaplygin gas have been made.\\
\end{abstract}

\section{Introduction}
\noindent 
A wide range of cosmological observations indicate
that the universe has entered a phase of accelerating expansion, which becomes one of the important
puzzles of the contemporary physics. Those observations include the type Ia supernovae (SnIa) standard candles \cite{hicken}, \cite{kowalski}, the angular location of the first peak in the CMB power spectrum \cite{komatsu} and baryon acoustic oscillations of the matter density power spectrum \cite{percival}. These evidences represent a great stimulus for theoretical work and originated the concept of Dark Energy. The Dark Energy (DE) Models attribute the observed accelerating expansion to an unknown energy component with negative pressure, which dominates the universe at recent cosmological times. The data can be accommodated with a high degree of accuracy, in the $\Lambda$CDM model which combines the baryons with conventional cold dark matter (CDM) candidates and the cosmological constant $\Lambda$, which accounts for the dark energy. However a non zero cosmological constant raises the coincidence problem (why the DM and DE are comparable today), and is plagued by fine tuning problems \cite{weinberg},\cite{peebles},\cite{padmana1}. Alternatively, dark energy may be described by different dynamical scalar field models with time-dependent equation of state, like quintessence \cite{RP}, \cite{wett}; string theory fundamental scalar known as tachyon \cite{pad}; K-essence models involving a generalized form of the kinetic energy \cite{stein3},\cite{chiba}; scalar field with negative kinetic term, which provides a solution known as phantom dark energy \cite{caldwell} (see \cite{copeland} for a review). An alternative description of DE may be given by perfect fluids with adequate equation of state, like Chaplygin gas \cite{bilic}, \cite{kamenshchick}. The above mentioned scalar field models assume that the DM and DE are of different nature, while in the Chaplygin gas the DM and DE appear as manifestations of the same single fluid at different cosmological epochs, showing a dust-like matter behavior at early times and behaving as cosmological constant at late times \cite{bilic}, \cite{kamenshchick}.\\
\noindent Another description of DE is provided by the scalar-tensor theories which contain a direct coupling of the scalar field to the curvature, with the advantage of giving a mechanism to evade the coincidence problem, and naturally allowing (in some cases) the crossing of the phantom barrier \cite{perivo}. These theories with different couplings to the curvature appear as low energy limit of several higher dimensional theories, and provide a possible approach to quantum gravity from a perturbative point of view \cite{donoghue}. 
A coupling between curvature and kinetic terms appears as part of the Weyl anomaly in $N=4$ conformal supergravity \cite{tseytlin, odintsov2}. 
A model with non-minimal derivative couplings was proposed in \cite{amendola2}, \cite{capozziello1}, \cite{capozziello2} in the context of inflationary cosmology, and recently, non-minimal derivative coupling of the Higgs field was considered in \cite{germani}, also as inflationary model. In \cite{caldwell1} a derivative coupling to Ricci tensor has been considered to study cosmological restrictions on the coupling parameter, and the role of this coupling during inflation. Some asymptotical solutions for a non-minimal kinetic coupling to scalar and Ricci curvatures were found in \cite{sushkov}, and quintessence and phantom cosmological scenarios with non-minimal derivative coupling have been studied in \cite{saridakis}. A scalar field with kinetic term coupled to a product of Einstein tensors has been considered in \cite{gao}. Non-minimal coupling of scalar fields (including kinetic terms) with modified $f(R)$ theories have been also considered to solve the DE problem in \cite{sergei}, \cite{sergei1}, \cite{allemandi}, \cite{sergei2}, \cite{sergei4}.\\
\noindent In this paper we consider an explicit coupling between the scalar field the kinetic term and the curvature \cite{granda, granda1, granda2}, as the source of DE and will establish the connection with the Chaplygin gas, by obtaining the solution of the field equations that reproduce the cosmological evolution as given by the perfect fluid obeying the equation of state of the Chaplygin gas \cite{kamenshchick}. The Chaplygin gas has attracted much attention in cosmology, as it allows to interpolate between a dust dominated phase of the evolution of the Universe in the past, and an accelerated one at recent time. On the other hand, as in the case of the Chaplygin gas, the scalar field with the non-minimal kinetic coupling to the curvature, has been shown to be successful in the description of the DM and DE without introducing separately the DM term (see \cite{granda1,granda2}), i.e. it describes the DM and DE as manifestations of a common scalar field at different epochs. For this reason, it would be interesting to analyze the connection of the scalar field with non-minimal kinetic couplings with Chaplygin gas, which will be the subject of study in the present paper. A potential and coupling function that may give a dynamical description to the Chaplygin gas, have been found. The generalized and modified version of the Chaplygin gas have been also considered.

\section{The Models and Field Equations}

The scalar field with kinetic couplings to curvature is given by \cite{granda1}:
\be\label{eq1}
\begin{aligned}
S=&\int d^{4}x\sqrt{-g}\Big[\frac{1}{16\pi G} R-\frac{1}{2}\partial_{\mu}\phi\partial^{\mu}\phi-\frac{1}{2} \xi R \left(F(\phi)\partial_{\mu}\phi\partial^{\mu}\phi\right) -\\ 
&\frac{1}{2} \eta R_{\mu\nu}\left(F(\phi)\partial^{\mu}\phi\partial^{\nu}\phi\right) - V(\phi)\Big].
\end{aligned}
\ee
And we will use the flat FRW background metric given by the line element
\be\label{eq1a}
ds^2=-dt^2+a^2(t)\left[dr^2+r^2\left(d\theta^2+\sin^2\theta d\phi^2\right)\right]
\ee
where $a$ is the scale parameter. The dimensionality of the coupling constants $\xi$ and $\eta$ depends on the type of function $F(\phi)$. Taking the variation of action (\ref{eq1}) with respect to the metric, we obtain a general expression of the form 
\be\label{eq2}
R_{\mu\nu}-\frac{1}{2}g_{\mu\nu}R=\kappa^2T_{\mu\nu}
\ee
where $\kappa^2=8\pi G$ and the tensor $T_{\mu\nu}$ represents the variation of the terms which depend on the scalar field $\phi$ and can be written as
\be\label{eq3}
T_{\mu\nu}=T_{\mu\nu}^{\phi}+T_{\mu\nu}^{\xi}+T_{\mu\nu}^{\eta}
\ee
where $T_{\mu\nu}^{\phi}$, $T_{\mu\nu}^{\xi}$, $T_{\mu\nu}^{\eta}$ correspond to the variations of the minimally coupled terms, the $\xi$ and the $\eta$ couplings respectively. Due  to the interaction between the scalar field and the curvature, the derived expressions for the density and pressure for the scalar field can be regarded as effective ones. From now on, and in order to simplify the equations (the field equations will contain only second order derivatives) we will use the restriction $\eta=-2\xi$, which is equivalent to a coupling of the kinetic term to the Einstein tensor $G_{\mu\nu}$ (see \cite{capozziello1}, \cite{capozziello2}). 
Evaluating the $00$ and $11$ components of the Eq. (\ref{eq2}) in the spatially-flat Friedmann-Robertson-Walker (FRW) background (\ref{eq1a}), it is obtained (with the Hubble parameter $H=\dot{a}/a$, and for homogeneous time-depending scalar field)
\be\label{eq4}
H^2=\frac{\kappa^2}{3}\left[\frac{1}{2}\dot{\phi}^2+9\xi H^2F(\phi)\dot{\phi}^2+V(\phi)\right]
\ee
and
\be\label{eq5}
-2\dot{H}-3H^2=\kappa^2\left[\frac{1}{2}\dot{\phi}^2-V(\phi)-\xi\left(3H^2+2\dot{H}\right)F(\phi)\dot{\phi}^2-2\xi H\left(2F(\phi)\dot{\phi}\ddot{\phi}+\frac{dF}{d\phi}\dot{\phi}^3\right)\right]
\ee
where ``dot'' represents the derivative with respect to the cosmological time $t$. Taking variation in (\ref{eq1}) with respect to the scalar field in the FRW background, gives the equation of motion as follows
\be\label{eq7}
\ddot{\phi}+3H\dot{\phi}+\frac{dV}{d\phi}+3\xi H^2\left(2F(\phi)\ddot{\phi}+\frac{dF}{d\phi}\dot{\phi}^2\right)
+18\xi H^3F(\phi)\dot{\phi}+12\xi H\dot{H}F(\phi)\dot{\phi}=0
\ee
where the first three terms describe the minimally coupled field.
In what follows we will study cosmological solutions to Eqs. (\ref{eq5}) and (\ref{eq7}) giving rise to accelerated expansion, and according to the cosmological scenario described by the Chapkygin gas solutions.\\
The Chaplygin gas is described by the following equation of state (EoS)
\be\label{eq8}
p=-\frac{A}{\rho}
\ee
where $p$ and $\rho$ are respectively the pressure and density, and $B$ is a positive constant. The continuity equation takes the form
\be\label{eq9}
\dot{\rho}+3H\left(\rho-\frac{A}{\rho}\right)=0
\ee
This equation can be easily integrated in the variable $x=\log a$, yielding
\be\label{eq10}
\rho=\left[A+B e^{-6x}\right]^{1/2}
\ee
which according to the Friedmann equation gives the following Hubble parameter
\be\label{eq11}
H^2=\frac{\kappa^2}{3}\left[A+B e^{-6x}\right]^{1/2}=\frac{\kappa^2}{3}\left[A+\frac{B}{a^6}\right]^{1/2}
\ee
This solution has the known advantages of describing the presureless matter dominance stage (epoch) of the universe at early times ($a<<1$, normalizing the current value of $a$ to $1$), and the future universe dominated by the cosmological constant, entering in a de Sitter phase at $a>>1$.  In the next section we will consider this solution to integrate the equations (\ref{eq4}) and (\ref{eq7}) with respect to the scalar field $\phi$ and the coupling $F$, and in this manner we will obtain a description of the Chaplygin gas cosmology in the frame of the scalar field with non-minimal kinetic coupling to curvature. Then, we will follow the same procedure with the generalized and modified Chaplygin gas models.

\section{Standard, generalized and modified Chaplygin gas solutions}

In order to integrate the Eqs. (\ref{eq4}) and (\ref{eq7}) for a given Hubble parameter, we should impose additional restrictions on the scalar field potential in order to consistently find the rest  of the variables, as follows (see \cite{granda1}).\\
In terms of the variable $x=\log a$, and defining the function $\theta(x)=\phi'^2$, the Eq. (\ref{eq7}) can be written as (after multiplying by $\dot{\phi}$)
\be\label{eq12}
\frac{1}{2}\frac{d}{dx}\left(H^2\theta\right)+3H^2\theta+\frac{dV}{dx}+9\xi H^2\frac{dH^2}{dx}F\theta+3\xi H^4\frac{d}{dx}(F\theta)+18\xi H^4F\theta=0
\ee
From Eq. (\ref{eq4}), changing to the variable $x$, we can write the product $F\phi'^2=F\theta$ as following 
\be\label{eq13}
F\theta=\frac{1}{3\xi\kappa^2 H^2}-\frac{\theta}{18\xi H^2}-\frac{V}{9\xi H^4}
\ee
taking the derivative of Eq. (\ref{eq13}) and replacing $F\theta$ and $d(F\theta)/dx$ into Eq. (\ref{eq12}), we arrive at the following equation involving $\theta$, $H$ and $V$
\be\label{eq14}
\begin{aligned}
&2H^4\frac{d\theta}{dx}+H^2\left(12H^2+\frac{dH^2}{dx}\right)\theta+4H^2\frac{dV}{dx}-2\left(6H^2+\frac{dH^2}{dx}\right)V\\
&+12\frac{H^2}{\kappa^2}\left(3H^2+\frac{dH^2}{dx}\right)=0
\end{aligned}
\ee
In this manner, we obtain a first order differential equation for the functions $\theta$, $H$ and $V$. In order to integrate the equation (\ref{eq14}), and thanks to the fact that the functions $\theta$ and $V$ are separated, we can impose a restriction on the scalar field potential given by the equation
\be\label{eq15}
2H^2\frac{dV}{dx}-\left(6H^2+\frac{dH^2}{dx}\right)V+\frac{6H^2}{\kappa^2}\left(3H^2+\frac{dH^2}{dx}\right)=0
\ee
which simplifies the Eq. (\ref{eq14}):
\be\label{eq16}
2H^2\frac{d\theta}{dx}+\left(12H^2+\frac{dH^2}{dx}\right)\theta=0
\ee
\noindent{\bf The Chaplygin gas solution}\\

In order to consistently solve the Eqs. (\ref{eq15}) and (\ref{eq16}), we propose the expression for the Hubble parameter $H^2$, given by that of the Chaplygin gas (\ref{eq11})
\be\label{eq17}
H^2=\frac{\kappa^2}{3}\left[A+B e^{-6x}\right]^{1/2}
\ee
Defining the scaled Hubble parameter $\tilde{H}=H/H_0$, Eq. (\ref{eq17}) can be written as
\be\label{eq18}
\tilde{H}^2=\left[\tilde{A}+\tilde{B} e^{-6x}\right]^{1/2}
\ee
where the parameters $\tilde{A}$ and $\tilde{B}$ are now dimensionless and are given by
\be\label{eq19}
\tilde{A}=\left(\frac{\kappa^2}{3H_0^2}\right)^2 A, \,\,\,\,\,\,\,\,\, \tilde{B}=\left(\frac{\kappa^2}{3H_0^2}\right)^2 B. 
\ee 
where  $\tilde{A}$ and $\tilde{B}$ satisfy the flatness condition (considering the Chaplygin gas dominance)
\be\label{eq19a}
\tilde{A}+\tilde{B}=1
\ee
Replacing  $\tilde{H}^2$ in (\ref{eq15}), and defining the dimensionless scalar potential $\tilde{V}=\kappa^2 V/H_0^2$, after changing the Eq. (\ref{eq15}) to the ``tilde'' variables and integration we obtain the scalar field potential 
\be\label{eq20}
\tilde{V}(x)=C e^{3x}\left(\tilde{A}+\tilde{B} e^{-6x}\right)^{1/4}-\frac{6\tilde{A}}{\tilde{B}} e^{6x}\left(\tilde{A}+\tilde{B} e^{-6x}\right)^{1/2}\hspace{0.1cm} _{2}F_1\left[1,\frac{1}{2},\frac{5}{4},-\frac{\tilde{A} e^{6x}}{\tilde{B}}\right]
\ee
where $C$ is the integration constant. Replacing $\tilde{H}^2$ in (\ref{eq16}) we get the following expression for $\theta$
\be\label{eq21}
\theta(x)=\phi'^2=\frac{\theta_0e^{-6x}}{\left(\tilde{A}+\tilde{B} e^{-6x}\right)^{1/4}},
\ee
where $\theta_0$ is the integration constant. Integrating the square root of this last equation, we obtain the scalar field as (considering the $(-)$ sign root)
\be\label{eq22}
\phi(x)=\frac{4\theta_0^{1/2} e^{-9x/4}}{9\tilde{B}^{1/8}}\hspace{0.1cm} _{2}F_1\left[\frac{1}{8},-\frac{3}{8},\frac{5}{8},-\frac{\tilde{A} e^{6x}}{\tilde{B}}\right]
\ee
Finally, the coupling function $F$ is found by replacing the Eqs. (\ref{eq18})-(\ref{eq21}) in the Friedmann Eq. (\ref{eq13}), giving the result
\be\label{eq23}
F(x)=\frac{e^{6x}g(x)^{-1/2}}{3\xi \kappa^2 H_0^2\theta_0}\Big[g(x)^{1/4}-\frac{\kappa^2\theta_0}{6}e^{-6x}
-\frac{C}{3} e^{3x}+\frac{2\tilde{A}}{\tilde{B}}e^{6x}g(x)^{1/4}\hspace{0.1cm} _{2}F_1\left[1,\frac{1}{2},\frac{5}{4},-\frac{\tilde{A} e^{6x}}{\tilde{B}}\right]\Big]
\ee
where $g(x)=\tilde{A}+\tilde{B} e^{-6x}$. Although we can not have an analytic expression for the potential in terms of the scalar field, we can illustrate the behavior of the potential as showed in fig.1, for a given set of parameters. The constant $C$ is selected so that the time variation of the gravitational coupling does not exceed the observational limits (see below). As can be seen from Fig. 1, for the selected values of the parameters, the potential is a monotonic decreasing function of the scalar field. The decreasing runaway behavior of the potential describing dark energy, is a key fact for a realistic cosmological model \cite{zlatev}, \cite{liddle}.
\begin{center}
\includegraphics [scale=0.7]{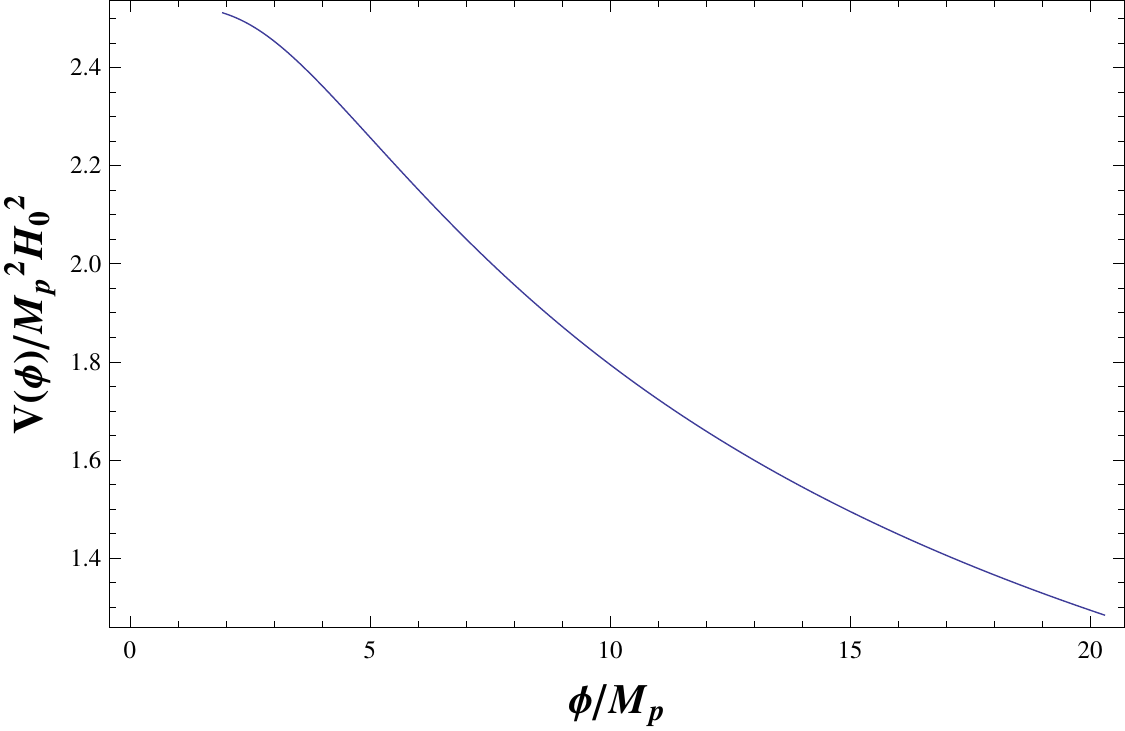}
\end{center}
\begin{center}
{Fig. 1 \it The evolution of the potential for the Chaplygin gas solution with the scalar field for $\tilde{A}=0.7, C=11$.}
\end{center}
\noindent{\bf The generalized Chaplygin gas solution}\\
A generalization of the Chaplygin gas has also been considered to describe the dark matter and dark energy. It's equation of state is given by
\be\label{eq24}
p_G=-\frac{A}{\rho_G^{\alpha}} 
\ee
where $\rho_G$ and $p_G$ are the energy density and pressure of the generalized Chaplygin gas, $A$ is a positive constant and $\alpha$ is considered to lie in the range $0<\alpha\leq 1$, which guarantees the stability and causality (see below) \cite{bilic}, \cite{kamenshchick}, \cite{bento}. Note that $\alpha=1$ corresponds to the original Chaplygin gas. The EoS (\ref{eq24}) has an equivalent field theory representation in a generalization of the Born-Infeld theory \cite{bento} (in the scalar field representation used in \cite{bento}, the Born-Infeld Lagrangian density is reproduced for $\alpha=1$). 
solving the continuity equation ($\dot{\rho}_G+3H(\rho_G+p_G)=0$), leads to the energy density in terms of $x=\log a$
\be\label{eq25}
\rho_G=\left(A+B e^{-3(\alpha+1)x}\right)^{\frac{1}{1+\alpha}}
\ee
where $B$ is a positive integration constant. This density for $\alpha>-1$ describes the matter dominated phase at $a<<1$ and at the limit $a>>1$ describes the de Sitter phase dominated by constant density. At the limit $\alpha\rightarrow 0$, Eq. (\ref{eq25}) reproduces the $\Lambda$CDM model. This model also captures the attention because of it's connection with string theory and supersymmetry: the EoS (\ref{eq24}) can be obtained in the Nambu-Goto action for $d$-branes moving in a $(d+2)$ dimensional space time \cite{hoppe}, and has supersymmetric generalization \cite{jackiw}. As a criteria to constraint the constant $\alpha$ we can use the sound speed for the fluid described by Eq. (\ref{eq24}), given by 
\be\label{eq25a}
c_s^2=\frac{dp_G}{d\rho_G}=\frac{\alpha A}{A+B e^{-3(1+\alpha)x}}=\frac{\alpha A}{A+B (1+z)^{3(1+\alpha)}}
\ee
where the last is written in terms of the redshift $z$ ($e^{-x}=(1+z)$). At future $z\rightarrow-1$, and as follows from (\ref{eq25a}) $c_s^2\rightarrow\alpha$. Therefore, stability requires $\alpha>0$, and causality requires $\alpha<1$ (i.e. respects the speed of light limit), limiting $\alpha$ to the interval $0\leq\alpha\leq 1$ (note that there is not such restriction on $\alpha$ at high redshift, during the matter dominated epoch). However some authors have considered the ``forbidden'' region $\alpha>1$ if the sound speed $c_s$ is treated as group velocity \cite{gorini1}, (see also \cite{yuko} for superluminal sound speed). According to some observational studies, for the case of the pure GCG the $\alpha<10^{-4}$ values are favored, which is very close to the $\Lambda$CDM limit \cite{bartolo}. This tight restriction may be increased if we consider additionally the barionic matter component ($\alpha<10^{-2}$) \cite{amendola1}, or adding cold dark matter and barionic matter components ($\alpha<0.2$) \cite{amendola3}. On the other hand, based on observations of the barionic power spectrum, the GCG plus barion matter is favored for $\alpha\geq 3$ \cite{gorini1}, \cite{yuko}. \\
Replacing the density (\ref{eq25}) in the Friedman equation gives the following scaled Hubble parameter
\be\label{eq26}
\tilde{H}^2=\left[\tilde{A}+\tilde{B} e^{-3(\alpha+1)x}\right]^{\frac{1}{1+\alpha}}
\ee
where the dimensionless $\tilde{A}$ and $\tilde{B}$ are given by
\be\label{eq27}
\tilde{A}=\left(\frac{\kappa^2}{3H_0^2}\right)^{1+\alpha} A, \,\,\,\,\,\,\,\,\, \tilde{B}=\left(\frac{\kappa^2}{3H_0^2}\right)^{1+\alpha} B 
\ee 
and satisfy the flatness condition (for pure generalized Chaplygin gas content)  $\tilde{A}+\tilde{B}=1$. 
Replacing (\ref{eq26}) in (\ref{eq15}) and solving the Eq. (\ref{eq15}) in ``tilde'' variables we obtain the following solution for the scalar potential
\be\label{eq28}
\begin{aligned}
\tilde{V}(x)=& C e^{3x}\left(\tilde{A}+\tilde{B} e^{-3(1+\alpha)x}\right)^{\frac{1}{2(1+\alpha)}}-\\
&\frac{6\tilde{A}}{(2\alpha-1)\tilde{B}} e^{3(1+\alpha)x}\left(\tilde{A}+\tilde{B} e^{-3(1+\alpha)x}\right)^{\frac{1}{1+\alpha}}\hspace{0.1cm} _{2}F_1\left[1,\frac{\alpha}{1+\alpha},2-\frac{3}{2(1+\alpha)},-\frac{\tilde{A} e^{3(1+\alpha)x}}{\tilde{B}}\right]
\end{aligned}
\ee
where $C$ is the integration constant. Solving (\ref{eq16}) with $\tilde{H}^2$ given by (\ref{eq26}), we find the expression for $\theta$
\be\label{eq29}
\theta(x)=\phi'^2=\frac{\theta_0e^{-6x}}{\left(\tilde{A}+\tilde{B} e^{-3(1+\alpha)x}\right)^{\frac{1}{2(1+\alpha)}}},
\ee
where $\theta_0$ is the integration constant. Integrating the square root of (\ref{eq29}) it follows
\be\label{eq30}
\phi(x)=\frac{4\theta_0^{1/2} e^{-9x/4}}{9\tilde{B}^{\frac{1}{4(1+\alpha)}}}\hspace{0.1cm} _{2}F_1\left[\frac{1}{4(1+\alpha)},-\frac{3}{4(1+\alpha)},\frac{1+4\alpha}{4(1+\alpha)},-\frac{\tilde{A} e^{6x}}{\tilde{B}}\right]
\ee
And the corresponding coupling function, as follows from (\ref{eq13}) and (\ref{eq26}-\ref{eq29}) is 
\be\label{eq31}
\begin{aligned}
F(x)=&\frac{e^{6x}g_{\alpha}(x)^{-1}}{3\xi\kappa^2 H_0^2\theta_0}\Big[g_{\alpha}(x)^{1/2}-\frac{\kappa^2\theta_0}{6}e^{-6x}-\frac{C}{3}e^{3x}+\\
&\frac{2\tilde{A}}{(2\alpha-1)\tilde{B}}e^{3(1+\alpha)x}g_{\alpha}(x)^{1/2}\hspace{0.1cm} _{2}F_1\left[1,\frac{\alpha}{1+\alpha},2-\frac{3}{2(1+\alpha)},-\frac{\tilde{A} e^{3(1+\alpha)x}}{\tilde{B}}\right]\Big]
\end{aligned}
\ee
where $g_{\alpha}(x)=\left(\tilde{A}+\tilde{B}e^{-3(1+\alpha)x}\right)^{\frac{1}{1+\alpha}}$. In all equations the dependence on the redshift $z$ or in the scale factor $a$ is obtained by replacing $e^{-x}=1+z=a^{-1}$. In fig.2 we plot the behavior of the potential in terms of the scalar field for two values of $\alpha$ corresponding to the ``physical'' region $0\leq\alpha\leq 1$, and the superluminal region $\alpha\sim 3$ which also accommodates in some astrophysical observations \cite{gorini1}, \cite{yuko}.  
\begin{center}
\includegraphics [scale=0.7]{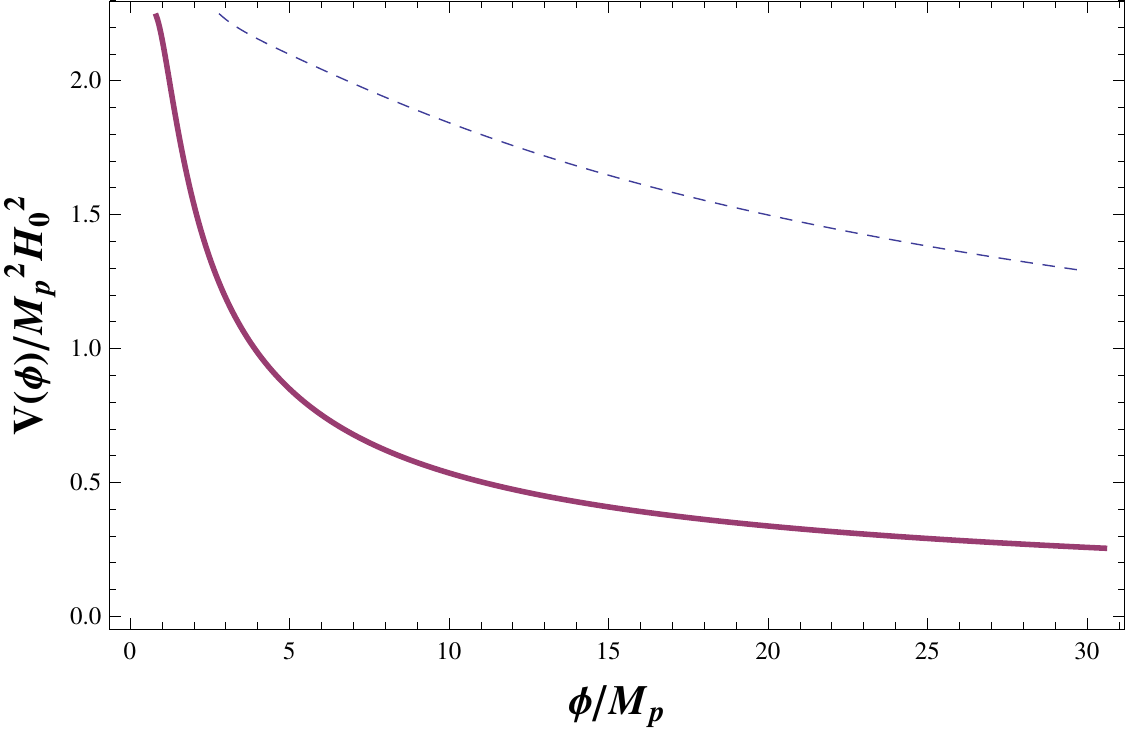}
\end{center}
\begin{center}
{Fig. 2 \it The potential for the GCG versus the scalar field for $\tilde{A}=0.5, \alpha=0.6, C=25.7$ (dashed), and $\tilde{A}=0.5, \alpha=3, C=3.1$. Note the decreasing behavior, which is an important characteristic for dark energy potentials}
\end{center}
The scalar field potentials for the non-minimally coupled scalar field, that reproduce the dynamics of the Chaplygin and generalized Chaplygin gas are decreasing functions of the scalar field. Note the runaway behavior for all curves, which are characteristic of well behaved dark energy potentials \cite{zlatev}, \cite{liddle}. The parameters are chosen in such a way that $V(\phi)$ is definite positive, at least for $z>-1$. For other choices of the parameters, it can be shown numerically that the curves can have one maximum. $C$ is used to accomplish the observational restrictions on the time variation of the gravitational coupling.\\
So, we have reconstructed the scalar model with kinetic couplings to curvature (\ref{eq1}), for a given Hubble parameter describing the CG and GCG cosmologies. In both cases we exploited the additional degree of freedom represented in the coupling function, to constraint the scalar potential in a way consistent with the Friedmann equations. A more general formulation of cosmological reconstruction method (in time and $x$ variables) for a number of modified gravities including scalar tensor theories, have been performed in \cite{sergei7}, \cite{sergei8}. In the scalar tensor theories considered in \cite{sergei7}, \cite{sergei8}, the scalar potential and couplings are reconstructed by using an apropriate redefinition of the scalar field, and giving the particular type of cosmological evolution encoded in $H$. The reconstruction was considered in the cosmological time and the e-folding variable $x$,  and concrete examples of accelerated late time cosmologies have been provided. In this works for the case of $f(R)$ gravity, the reconstruction was achieved by introducing an auxiliary scalar field.

\noindent{\bf The modified Chaplygin gas solution}\\
The modified Chaplygin gas (MCG) is defined for the equation of state
\be\label{eq32}
p=B\rho-\frac{A}{\rho^{\alpha}}
\ee
A generalized version of an equation of state that includes CG, GCG and MCG have been considered in \cite{sergei5} and \cite{sergei6}. Integrating the continuity equation gives the following energy density
\be\label{eq33}
\rho=\left(\frac{A}{1+B}+Ce^{-\beta x}\right)^{\frac{1}{1+\alpha}}=\left(\frac{A}{B+1}+\frac{C}{a^{\beta}}\right)^{\frac{1}{1+\alpha}}
\ee
where $\beta=3(1+B)(1+\alpha)$ and $C$ is the integration constant. This equation has the same functional dependence on $x$ or $a$, as the case of the GCG, except that additional limits can be obtained due to additional factor $(1+B)$ in the exponent of Eq. (\ref{eq33}), with respect to the corresponding one in Eq. (\ref{eq25}). 
Therefore, the resulting potential, scalar field and coupling function corresponding to the MCG shows the same dependence as given in Eqs. (\ref{eq28}-\ref{eq31}), and it can be shown that all the equations of the MCG, at $B=0$ become the corresponding equations for the GCG. Additionally to the GCG, the MCG model reproduces the radiation dominated phase of the universe at high redshift, as can be seen from (\ref{eq33}) for $B=1/3$ and $a<<1$ (neglecting the constant term). In general the MCG reproduces the barotopic fluid with constant equation of state ($\rho\sim C^{1/(1+\alpha)}a^{-3(1+B)}$) at earlier epochs when the constant term may be neglected. Resumming, the non-minimally coupled scalar field also reproduces the dynamics of the modified Chaplygin gas cosmology, with well behaved scalar field, potential and coupling function.\\

\noindent{\bf The time variation of the gravitational coupling}\\
We can meet the constraints on the current value and the time variation of the gravitational coupling \cite{uzan1}, by appropriately defining or constraining the constants $C$ and $\theta_0$. The effective gravitational coupling from (\ref{eq4}) is given by
\be\label{eqg1}
G_{eff}=\frac{G}{1-3\xi\kappa^2FH^2\theta}
\ee
where we used $\kappa^2=8\pi G$ and $\dot{\phi}^2=H^2\phi'2=H^2\theta$. In terms of $x$, the time variation of the gravitational coupling can be written as
\be\label{eqg2}
\frac{\dot{G}_{eff}}{G_{eff}}=\frac{3\xi\kappa^2}{1-3\xi\kappa^2FH^2\theta}\frac{d}{dx}(FH^2\theta)H.
\ee
\noindent Replacing the product $FH^2\theta$ from Eq. (\ref{eq13}), and evaluating at the present time ($x=0$), the Eq. (\ref{eqg2}) can be written as
\be\label{eqg3}
\frac{\dot{G}_{eff}}{G_{eff}}\Big|_{x=0}=\frac{3f(C,\theta_0)}{1-3g(C,\theta_0)}H_0
\ee
where $f(C,\theta_0)=\xi\kappa^2 d(FH^2\theta)/dx$, $g(C,\theta)=\xi\kappa^3 FH^2\theta$ valuated at $x=0$ and the parameters of the model appearing in $\tilde{H}^2$ have been fixed, so that the resulting expression (\ref{eqg3}) depends on the constants of integration $C$ and $\theta_0$.
We can meet the restrictions imposed by the current observations on the value and the time variation of the gravitational coupling \cite{uzan1}, if $f(C,\theta)$ and $g(C,\theta)$ satisfy the constraints: $f(C,\theta)\approx 0$ and $g(C,\theta)\approx 0$ (these restrictions are actually softer: $f(C,\theta_0)\leq10^{-1}$ and $g(C,\theta_0)\leq10^{-5}$).
Thus, for the Chaplygin gas solution given by (\ref{eq17}), (\ref{eq20}-\ref{eq23}) with $\tilde{A}=0.7$ ($\tilde{B}=0.3$), under above restrictions we found $C\approx 11$ and $\kappa^2\theta_0\approx 0.9$, which give the potential plotted in Fig. 1. Following the same procedure for the generalized Chaplygin gas, we found $C\approx 25.7$, $\kappa^2\theta_0\approx 1.5$  (for $\tilde{A}=0.5$ ($\tilde{B}=0.5$), $\alpha=0.6$) and  $C\approx 3.1$, $\kappa^2\theta_0\approx 1.5$  for the same values of  ($\tilde{A}$, $\tilde{B}$) and $\alpha=3$, with the respective potentials plotted in fig. 2.

\section{Discussion}

The cosmological implications of the Chaplygin gas model have been intensively investigated in recent literature. 
We considered the model of scalar field with kinetic terms coupled non-minimally to the scalar field and to the curvature, to give a dynamical description of the Chaplygin gas model of dark energy.  We have found analytical expressions for the reconstructed scalar field and potentials that describe the standard, the generalized and modified Chaplygin gas models of dark energy dark mater unification. Thanks to the presence of the coupling function $F(\phi)$, we could impose a restriction on the potential through Eq. (\ref{eq15}), which allowed us to find the solutions that lead to dynamical description of the Chaplygin gas cosmology. The results show that the obtained potentials $V(\phi)$ decrease with the evolving
scalar field $\phi$. From Eqs. (\ref{eq20},\ref{eq22}) it follows that $\phi$ is an increasing function of the redshift and $V$ is a decreasing function of the redshift, which means that the scalar potential is a decreasing function of the scalar field. In Fig 1. we show the $V(\phi)$ dependence for the standard Chaplygin gas in the redshift interval $[0,2]$. The same behavior follows from Eqs. (\ref{eq28},\ref{eq30}) for the GCG as shown in Fig. 2. The runaway behavior of all the potentials is a desired property of dark energy potentials as the relevance of the scalar potential increases at the present epoch (low redshift), as showing in Figs. 1 and 2. 
Although the problem of perturbations has not been considered here, it should be noted that, despite of the equivalence found between the specific reconstructed scalar model and the Chaplygin gas, under perturbations of the scalar field (which induce additional perturbations of the metric through the kinetic coupling) it may be expected that this equivalence could not go beyond perturbative corrections.\\
In order to satisfy the current restrictions on the Newtonian coupling, we can use the freedom in the integrations constants $C$ and $\theta_0$, in order to control the actual value of $G$ and it's time variation. These conditions can be satisfied by imposing the inequalities $f(C,\theta_0)\leq10^{-1}$ and $g(C,\theta_0)\leq10^{-5}$.\\
\noindent The above results show that the scalar field model with derivative couplings to curvature considered here, provide a dynamical scenario to describe the Chaplygin gas cosmology. The wide variety of phenomenologically acceptable solutions \cite{granda,granda1,granda2}, support the capability of this model to explain the current status of the accelerated expansion of the universe, through different cosmological scenarios. 

\section*{Acknowledgments}
This work was partially supported by the SENECA foundation (Spain) under the program PCTRM 2007-2010 and the Spanish Ministry of Science 
and Education grant FPU AP2008-00919 (JJFM). LNG thanks the CERN TH-Division for kind hospitality.

\end{document}